\documentclass[11pt]{article}
\usepackage{moriond,epsfig}

\def\be{\begin{equation}}
\def\ee{\end{equation}}
\def\bea{\begin{eqnarray}}
\def\eea{\end{eqnarray}}

\begin{document}
\vspace*{4cm}
\title{REHEATING IN AN EARLY SUPERSYMMETRIC UNIVERSE}

\author{ANNA KAMI\'{N}SKA AND PAWE{\L} PACHO{\L}EK}

\address{Institute of Theoretical Physics, Faculty of Physics, \\
University of Warsaw, Ho{\.z}a 69, Warsaw, Poland}

\maketitle

\abstracts{
Motivated by a recent discussion about the role of flat directions, a typical feature of supersymmetric models, in the process of particle production in the early universe a consistent model of inflation and preheating in supergravity with MSSM fields has been built. It is based on a model proposed by M. Kawasaki, M. Yamaguchi and T. Yanagida. In the inflationary stage, the flat directions acquire large vacuum expectation values (VEVs) without spoiling the background of slow-roll, high-scale inflation consistent with the latest WMAP5 observational data. In the stage of particle production, naturally following inflation, the role of flat direction large VEVs depends strongly on effects connected with the supergravity framework and non-renormalizable terms in the superpotential, which have been neglected so far in the literature. Such effects turn out to be very important, changing the previous picture of preheating in the presence of large flat direction VEVs by allowing for efficient preheating from the inflaton.}

\section{Introduction}

In order to properly describe inflation and particle production one has to consider the underlying theory of particles and interactions. Supersymmetry is one of the most promising extensions of the Standard Model (SM). One of the typical features of supersymmetric extensions of the SM is the presence of flat directions \cite{gherghetta} - directions in field space, along which the scalar potential identically vanishes in the limit of unbroken global supersymmetry. Due to large quantum fluctuations or the classical evolution of fields during inflation flat directions can easily acquire large VEVs. Therefore, there is a natural question about the role of such large VEVs in the process of particle production.\newline
It was postulated \cite{mazumdar} that large flat direction VEVs influence the process of particle production by blocking preheating from the inflaton - the phase of rapid, non-perturbative inflaton decay. A simple toy model was proposed
\begin{equation}
V\supset{}A\varphi^{2}\chi^{2}+B{}m\varphi\chi^{2}+C\alpha^{2}\chi^{2},
\label{spi}
\end{equation}
where $\varphi$ is the inflaton field, $\alpha$ parameterizes the flat direction and $\chi$ represents the inflaton decay products (in this model a direction in Higgs fields has been considered). Then, after mode decomposition of the field $\chi$, the energy of the mode with momentum $k$ is given by:
\begin{equation}
\omega_{k}^{2}=k^{2}+2A\left\langle\varphi\right\rangle^{2}+2B{}m\left\langle\varphi\right\rangle+2C\left\langle\alpha\right\rangle^{2}.
\end{equation}
In general, non-adiabatic production of particles $\chi_{k}$ is efficient only when $\omega_{k}$ changes non-adiabatically
\begin{equation}
|\tau|\equiv\left|\frac{\dot{\omega}}{\omega^{2}}\right|>1\leftrightarrow{}preheating,
\label{adiabat}
\end{equation}
where the adiabaticity parameter $\tau$ is introduced. In the simplest model of inflation the epoch of slow-roll of the inflaton VEV usually ends with oscillations of the inflaton VEV around the minimum of its potential which is connected with particle production. During classical preheating $\omega_{k}$ is dominated by the inflaton VEV and changes non-adiabatically due to inflaton oscillations. In the presence of flat directions however, $\omega_{k}$ could be dominated by the large VEV of the flat direction. If this VEV changes very slowly in comparison with the evolution of the inflaton VEV, non-perturbative production of $\chi$ particles is effectively blocked.\newline
However blocking of preheating from the inflaton does not occur when non-perturbative production of particles from the flat direction itself is possible \cite{olive}. Then the initially large VEV of the flat direction decreases rapidly, unblocking preheating from the inflaton. This led to a discussion \cite{allahverdi} \cite{basboll1} \cite{basboll} \cite{olive1} about whether non-perturbative decay of flat directions and preheating from the inflaton is possible. The discussion did not consider any specific model of inflation and did not propose any model of acquiring large VEVs by flat directions.\newline
The goal of our work \cite{kaminska} is to construct a consistent model of inflation and particle production in a realistic supersymmetric extension of the SM, and consider in this specific model the behavior of MSSM flat directions. Therefore a realistic chaotic inflation model with two representative flat directions is constructed. In this specific model the process of acquiring large VEVs by flat directions and the impact of such VEVs on particle production is studied, including supergravity effects which have been neglected so far in literature. We find that these effects strongly influence the process of particle production by introducing efficient channels of non-perturbative particle production both from the flat direction and the inflaton. As a result the originally large flat direction VEVs are diminished, preheating from the inflaton is allowed and the energy density of the Universe is dominated by the inflaton decay products.

\section{The model}

The inflaton sector \cite{kawasaki} consists of two chiral superfields
\begin{equation}
\Phi=(\eta+i\varphi)/\sqrt{2},\ \ \ \ \ \ X=xe^{i\beta},
\end{equation}
where field $\varphi$ is the inflaton field. The choice of the K\"{a}hler potential and the superpotential
\begin{equation}
K\supset\frac{1}{2}(\Phi+\Phi^{\dag})^{2}+XX^{\dag},\ \ \ \ \ \ W\supset mX\Phi
\label{Kh}
\end{equation}
solves the eta-problem and provides the chaotic-type inflaton potential $V=1/2\;m^{2}\varphi^{2}$ during inflaton domination. Two MSSM-flat directions $H_{u}H_{d}$ and $udd$ parameterized by
\begin{equation}
H_{d}=\frac{1}{\sqrt{2}}\left(\begin{array}{cc} 
      \chi \\ 0 \end{array}\right),\ \ H_{u}=\frac{1}{\sqrt{2}}\left(\begin{array}{cc} 
       0 \\\chi \end{array}\right),\ \ \chi=ce^{i\kappa},\ \ u^{\beta}_{i}=d^{\gamma}_{j}=d^{\delta}_{k}=\frac{1}{\sqrt{3}}\alpha,\ \ \alpha=\rho e^{i\sigma}
\end{equation}
are introduced in the model in the following way
\begin{equation}
K\supset\left(1+\frac{a}{M^{2}_{Pl}}XX^{\dag}\right)\left(H_{u}H_{u}^{\dag}+H_{d}H_{d}^{\dag}+u_{i}u_{i}^{\dag}+d_{j}d_{j}^{\dag}+d_{k}d_{k}^{\dag}\right)
\label{K_NM}
\end{equation}
\begin{equation}
W\supset2hXH_{u}H_{d}+W_{MSSM}+\frac{\lambda_{\chi}}{M_{Pl}}\left(H_{u}H_{d}\right)^{2}+\frac{3\sqrt{3}\lambda_{\alpha}}{M_{Pl}}\left(u_{i}d_{j}d_{k}\nu_{R}\right).
\label{W}
\end{equation}
The non-minimal K\"{a}hler coupling scaled by $a$ creates a potential for the flat direction which has a minimum at large flat direction VEVs. This allows acquiring large flat directions VEVs by classical evolution during inflation. Parameter $h$ scales the only renormalizable coupling between the inflaton sector and the observable sector while $\lambda_{\chi}$ and $\lambda_{\alpha}$ scale non-renormalizable terms for flat directions in the superpotential and are used to control the position of the minima of the scalar potential.

\section{Classical evolution of fields}

\begin{figure}[h!]
\begin{minipage}[b]{0.47\linewidth}
\centering
\includegraphics[width=6 cm]{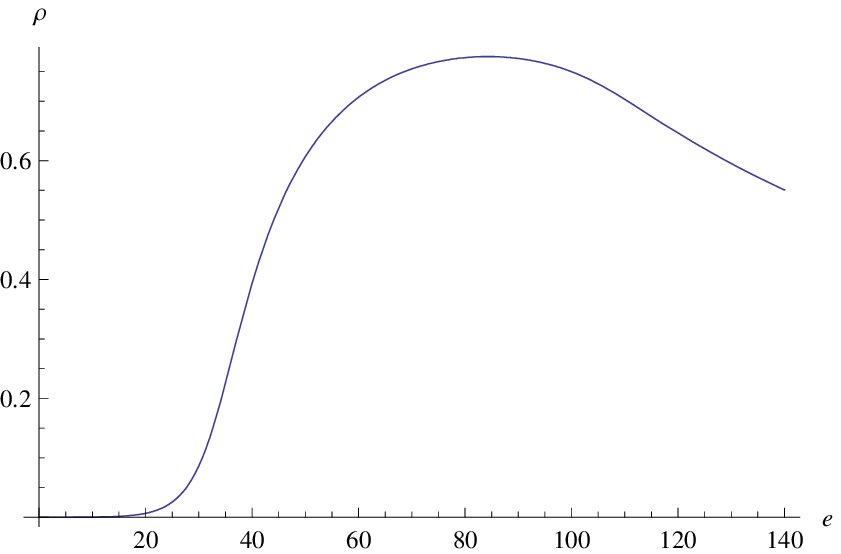}
\caption{Evolution of the field $\rho$ during inflation for $\lambda_{\alpha}=10^{-7}$ and $\lambda_{\chi}=1$. Values on vertical axes are expressed in Planck masses and time on horizontal axes is expressed in the approximate number of e-folds of inflation.}
\label{f1}
\end{minipage}
\hspace{0.5cm}
\begin{minipage}[b]{0.47\linewidth}
\centering
\includegraphics[width=6 cm]{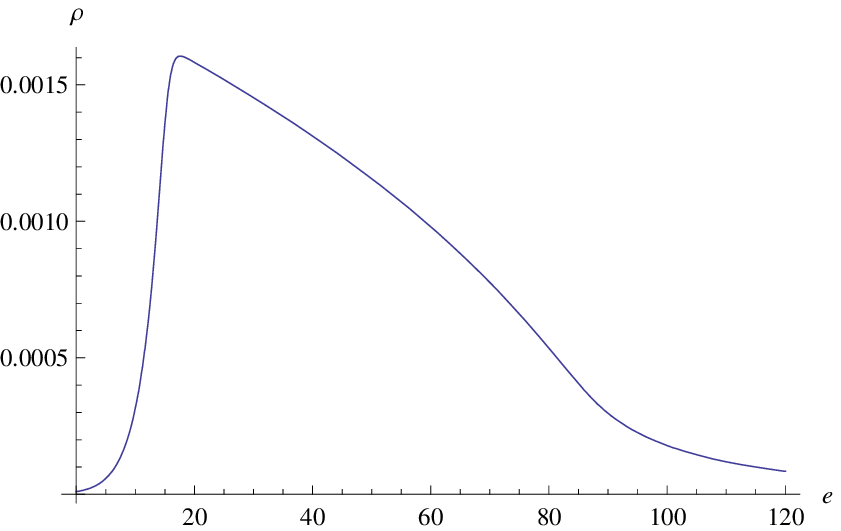}
\caption{Evolution of the field $\rho$ during inflation for $\lambda_{\alpha}=1$ and $\lambda_{\chi}=1$. Values on vertical axes are expressed in Planck masses and time on horizontal axes is expressed in the approximate number of e-folds of inflation. Evolution of the field $c$ is similar.}
\label{f2}
\end{minipage}
\end{figure}

The classical evolution of fields during inflation is obtained numerically and shows that acquiring large flat directions VEVs during inflation is possible even starting with small initial values $\delta\rho,\;\delta{}c\sim{}H\sim10^{-5}M_{Pl}$ corresponding to the average size of quantum fluctuations during inflation. For a single flat direction values $\sim0.8M_{Pl}$ can be acquired during inflation, as can be seen in Fig. \ref{f1}. Both $H_{u}H_{d}$ and $udd$ cannot acquire simultaneously such large VEVs as in the previous case because they are non-independent due to Yukawa couplings. With our choice of couplings they can acquire simultaneously VEVs $\sim10^{-3}M_{Pl}$ (Fig. \ref{f2}), which are still large in comparison with the Hubble parameter.

\section{Excitations around VEVs and particle production}

\begin{figure}[h!]
\begin{minipage}[b]{0.47\linewidth}
\centering
\includegraphics[width=6 cm]{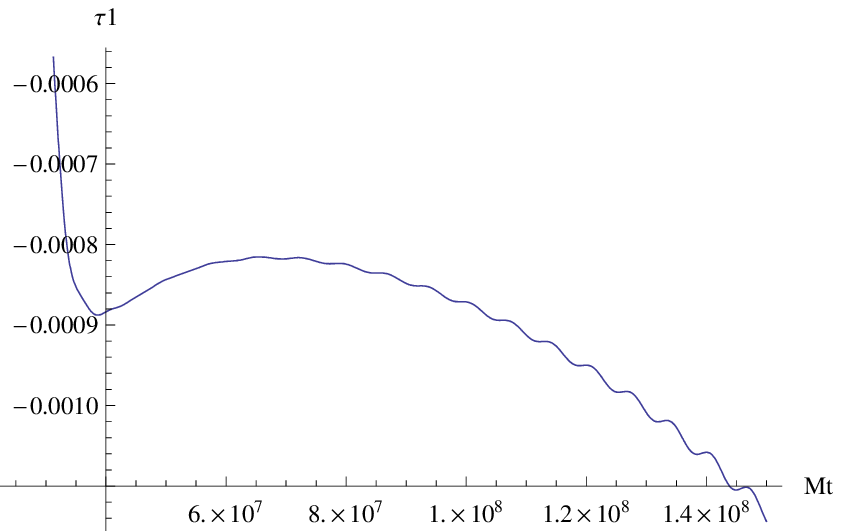}
\caption{Evolution of the adiabaticity parameter related to a heavy eigenvalue dominated by the flat direction VEV.}
\label{f3}
\end{minipage}
\hspace{0.5cm}
\begin{minipage}[b]{0.47\linewidth}
\centering
\includegraphics[width=6 cm]{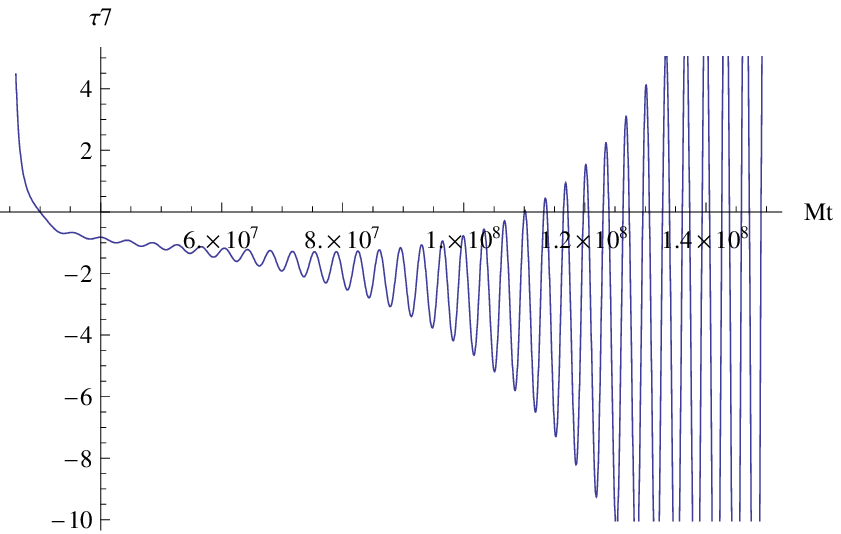}
\caption{Evolution of the adiabaticity parameter related to a light eigenvalue related to the flat direction VEV.}
\label{f4}
\end{minipage}
\end{figure}

In order to study particle production excitations around classically evolving VEVs are considered during inflaton oscillations. In particular the possibility of preheating into excitations of fields belonging to multiplets related to both $H_{u}H_{d}$ and $udd$ directions is studied. For fields with non-zero VEVs a non-linear parameterization of excitations is adopted
\begin{equation}
field=\left(\frac{|VEV|}{\sqrt{n}}+\frac{\xi_{1}}{\sqrt{2}}\right)e^{i\left(arg\left(VEV\right)+\frac{\xi_{2}}{\sqrt{2}|VEV|}\right)}.
\end{equation}
where $n=2$ for Higgs doublets and $n=3$ for squark triplets. After going to the unitary gauge the evolution of the mass matrix eigenvalues and eigenvectors is considered in order to check the possibility of non-perturbative particle production. As expected several channels of preheating are initially blocked due to the large, slowly evolving flat direction VEVs which induce heavy, adiabatically evolving mass matrix eigenvalues (Fig. \ref{f3}). However also light mass matrix eigenvalues appear. It can be understood remembering that in global SUSY without non-renormalizable terms each complex field parameterizing a flat direction is connected with two mass matrix eigenvalues which are equal zero, hence they do not contribute to non-perturbative particle production. Introducing a scalar potential dependent on the flat directions by supergravity effects makes these eigenvalues naturally light and strongly dependent on the quickly evolving inflaton VEV, which leads to the existence of very efficient channels of preheating from the inflaton (Fig. \ref{f4}). Non-perturbative particle production from flat directions into these channels is also allowed which leads to the rapid decrease of the flat direction VEVs and unblocks all other channels of preheating.

\section{Conclusions}

Achieving large flat direction VEVs through classical evolution during inflation is natural in a supergravity framework with non-minimal K\"{a}hler potential. Such large VEVs can block preheating from the inflaton into certain channels. However supergravity effects and non-renormalizable terms, which create a potential for the flat direction, are a source of light, rapidly changing eigenvalues of the mass matrix. They allow the non-perturbative production of particles from the flat direction and preheating from the inflaton. Thus non-perturbative particle production from the inflaton is likely to remain the source of preheating even in the initial presence of large flat direction VEVs.

\section*{Acknowledgments}

The authors would like to thank very much prof. Stefan Pokorski, prof. Keith Olive and prof. Marco Peloso for all their help with this work.

\end{document}